\documentclass[10pt, conference]{IEEEtran}
\usepackage{booktabs}
\usepackage{multirow}
\usepackage{cite}

\newtheorem{theorem}{Theorem}[section]
\newcommand{\bTheorem}{ \begin{theorem}  }
\newcommand{\eTheorem}{ \end{theorem}    }

\newtheorem{Definition}{Definition}[section]
\newcommand{\bDefinition}{ \begin{Definition} }
\newcommand{\eDefinition}{ \end{Definition} }
\newcommand{\bDef}{ \begin{Definition} }
\newcommand{\eDef}{ \end{Definition} }

\ifCLASSINFOpdf
\else
  \usepackage[dvips]{graphicx}
\fi
\usepackage[cmex10]{amsmath}

\begin{document}
%
\title{On the Interaction of Adaptive Video Streaming with Content-Centric Networking}



\author{
  \IEEEauthorblockN{Reinhard Grandl\IEEEauthorrefmark{1}, Kai Su\IEEEauthorrefmark{2} and Cedric Westphal\IEEEauthorrefmark{3}}
  \IEEEauthorblockA{
    \IEEEauthorrefmark{1}Institute of Networked and Embedded Systems, Alpen-Adria Universit\"at Klagenfurt, Austria
    }
  \IEEEauthorblockA{
    \IEEEauthorrefmark{2}WINLAB, Rutgers University, North Brunswick, NJ
    }
  \IEEEauthorblockA{
    \IEEEauthorrefmark{3}Huawei Innovation Center, Santa Clara, CA \& Computer Engineering Dept, University of California, Santa Cruz, CA\\
    Email: \IEEEauthorrefmark{1}r.grandl@edu.uni-klu.ac.at, \IEEEauthorrefmark{2}kais@winlab.rutgers.edu, \IEEEauthorrefmark{3}
    cedric.westphal@huawei.com}
}

\maketitle

\begin{abstract}
Two main trends in today's internet are of major interest for video streaming services: most content delivery platforms coincide towards using adaptive video streaming over HTTP and new network architectures allowing caching at intermediate points within the network. We investigate one of the most popular streaming service in terms of rate adaptation and opportunistic caching. Our experimental study shows that the streaming client's rate selection trajectory, i.e., the set of selected segments of varied bit rates which constitute a complete video, is not repetitive across separate downloads. Also, the involvement of caching could lead to frequent alternation between cache and server when serving back client's requests for video segments. These observations warrant cautions for rate adaption algorithm design and trigger our analysis to characterize the performance of in-network caching for HTTP streaming. Our analytic results show: (i) a significant degradation of cache hit rate for adaptive streaming, with a typical file popularity distribution in nowadays internet; (ii) as a result of the (usually) higher throughput at the client-cache connection compared to client-server one, cache-server oscillations due to misjudgments of the rate adaptation algorithm occur. Finally, we introduce DASH-INC, a framework for improved video streaming in caching networks including transcoding and multiple throughput estimation.
\end{abstract}

\section{Introduction}
\label{sec:introduction}

Most of the traffic over the Internet has shifted to video content, and it is predicted the fraction of Internet traffic consumed by video will reach 90\% in 2017~\cite{Cisco2013}. Currently, the video streaming service Netflix consumes a third of all web traffic in the evening~\cite{Kafka2013} and YouTube 17\%. These two services combine for half of the Internet traffic during prime time, and other services (Hulu, Amazon, HBO GO) are increasing in popularity. Since the current growth rate for the bandwidth demand is difficult to sustain for the network operators, new architectures have been proposed to cache the traffic within the network, and alleviate the bandwidth bottleneck.

Information-centric networks (ICN)~\cite{Ahlgren2012}, or content-centric networks (CCN)~\cite{Jacobson2009} are proposals for such new architectures which allow the network to cache data by attaching storage to routers. The routers thus become both forwarding elements and caching elements. In some such proposals, such as NDN~\cite{Zhang2010}, the caching is performed opportunistically. New objects are inserted in the cache as they reach the router, and older items are expunged from the cache to make room when the cache is full, according to a caching policy such as, say, the least recently used (LRU) policy.

Video streaming currently uses multiple technologies and protocols, but to allow inter-platform convergence, most content delivery platform seem to converge towards DASH (Dynamic Adaptive Streaming over HTTP) as the method to stream videos~\cite{Stockhammer2011}. There are multiple types of adaptive video formats, but they all use similar mechanisms: the video client starts downloading some segments from the server, and estimates from this download the highest rate that the network connection can support. 

With adaptive streaming, when a client A requests the video V from the server, it is as a series of segments indexed by time, and for each time index, one of multiple coding schemes (video playback rates) that are available. Client A views segments of the form $V_t^r$ with time $t = 1,\ldots,n$ and rate $r \in \{r_1,r_2,\ldots,r_K\}$. A video stream therefore becomes a trajectory over the (time,rate) space. When client B requests the same video V, it will see a series of segments that may or may not overlap with that of client A. Further, adaptive video streaming creates a feed-back loop between the client and the server. From the download of the previous segment, the client infers the channel conditions, and requests the proper rate. By inserting copies at points on the path, this channel is perturbed.

We might see the following effect: client A requests video V from the server S based upon the S-A conditions. This stream is cached as a (time,rate) trajectory at router R. For a subsequent request for V, some segments may be served directly from R to A. Client A thus computes the network performance based upon the delivery of the segments from R. Since it is much better than from S, the client increases its requested playback rate. Because the download trajectory is now different from the previous transfer from S cached at R, the new playback rate is not available at R. Thus the request is propagated back to S. However, this rate is not sustainable on the S-A channel, and the rate reverts back to the previously achievable rate, back to the trajectory cached at R. So the video stream may oscillate between segments at the cache and at the server.

We thus have two trends colliding: a shift from the point of view of the consumption of video towards dynamic adaptive video streaming such as DASH; and a shift from the point of view of the network architecture towards caching transparently (from the point of view of the user) at intermediate points within the network. We study in this paper the interaction of these two trends. In particular, we conduct {\em an experimental study} to consider a model of the segments that are requested by a video streaming client. We study the impact of using an adaptive video rate on the in-network caching performance. We {\em empirically observe the oscillations} between downloads from a cache and from the server created by rate adaptation. We {\em propose an analytic framework} to evaluate the performance degradation caused by these interactions, and {\em propose DASH-INC}, a rate adaptation mechanism that takes into account in-network caching. 

The rest of the paper is organized as follows: the next section contains references to related work. In section~\ref{sec:empiricalObservations} we describe our empirical observations in terms of: (i) rate adaptation of the Netflix video streaming client; (ii) cache-server oscillation in our ICN testbed using DASH with a VLC client. Section~\ref{sec:characterizing} analyzes the performance of in-network caching based upon our observation. We investigate throughput oscillations observed at a streaming client and induced by the presence of a cache in section~\ref{sec:cacheOscillation}. Section~\ref{sec:guidelines} discusses how to cache adaptive video streams and describes a way to overcome cache oscillations. Finally, in section~\ref{sec:conclusions}, we discuss and conclude the paper.

\section{Related Work \& Background}
\label{sec:relatedWork}

Much recent work has considered different aspects of video streaming and rate adaptation. Adhikari~\textit{et al.} performed measurements to uncover and evaluate Netflix~\cite{Adhikari2012}. In~\cite{Huang2012} the authors did a performance study of dynamic rate adaptation for several streaming services, including Netflix, Hulu and Vudu. They identified a \textit{downward spiral effect} - a dramatic anomalous drop in the video playback rate. This illustrates the interactions of the rate adaptation mechanism with the underlying network. As we will see, similar interactions of the streaming rate and the network layer are exacerbated by inserting in-network caching. The interaction of video streaming with TCP where studied (and to some extent, solved) in~\cite{Ghobadi2012}.

ICNs~\cite{Ahlgren2012} (say CCN~\cite{Jacobson2009} or DONA~\cite{Koponen2007}) were proposed to facilitate the distribution of content, and of video in particular. In ICNs, caching becomes part of the network service where all nodes potentially have caches (on-path caching). A request for a specific object can be satisfied by any node holding a copy in its cache.

Recently, Fayazbakhsh~\textit{et al.} discussed an incremental deployment of an ICN~\cite{Fayazbakhsh2013}, where caching happens only at the client's edge. This is the simple abstracted model we adopt in our paper: we consider a client connected to a local cache which can deliver some content and some enhanced, accelerated services, or can forward content requests onward to an origin server. This is similar to current HTTP proxies and caches~\cite{Wolman1999}, and our results have applicability beyond ICNs to such systems of web proxies. \cite{Chanda2013a,Chanda2013} specified an architecture where this edge cache is distributed over an access network domain, and managed by a modified content-aware SDN controller. This controller allows content routing over IP network for HTTP requests. We leverage this architecture for our evaluation.

Adaptive video streaming has been the object of much work. The purpose of adapting the video rate is to efficiently use the available network resource to deliver good QoE (quality of experience) at the client, but to not use more resource than can be supported by the network, so as to limit the work at the server. Layered video coding such as~\cite{Schwarz2007} split the video streams into a base stream and some additional layers with extra information which can be transmitted if the network supports it. FlexCast~\cite{Aditya2011} is a proposal to make the video quality proportional to the wireless channel conditions, but requires the server to spray many more packets than will hit the target. Several video streaming mechanisms (such as Apple's HTTP Live Streaming~\cite{httpLiveStreaming}, Microsoft's Smooth Streaming or Adobe's HTTP Dynamic Streaming) use similar concepts to DASH~\cite{Stoekhammer2011}, by having a stateless server sending only as much traffic as the client can receive, as estimated by the client. DASH seems to be the mechanism towards which video distribution over the Internet is converging. For this reason, we focus our study on the interaction of a DASH-like mechanism with an ICN architecture.

DASH works as follows: the client first requests a manifest file from the server that includes the media description presentation (MPD). This manifest describes the representation of the video stream as well as potential URLs to download the video. The video stream is a set of time segments, each one of which available in multiple encoding rates. The manifest describes which ones are available at the server. The client, upon receiving the manifest, can start requesting segments. DASH delivers just-in-time, so that only a limited quantity of data is dumped if a user cancels a video. The client starts requesting segments at a low rate, and from the time it takes to receive the segment, it assesses the network conditions. It can then increase and adapt the rate to the network conditions.

Note that in-network caching is different from CDN: the server for the video stream can be (and is in most services) a CDN node. Per the ICN principles, we consider caching in the network {\em in addition} to the use of a content distribution network. The next section studies the behavior of DASH in the wild, through an empirical study.

\section{Empirical Observations}
\label{sec:empiricalObservations}

\subsection{Correlation Between Subsequent Streams}

We are interested in the behavior of adaptive video streaming and how such mechanism would be impacted by the presence of in-network caching. If consecutive requests follow different rates, then in-network caching would yield little benefit. We conducted some experiments to assess the variability of the rate selection for an adaptive streaming client. We selected the Netflix client, as it consumes a large fraction of the overall web traffic (over 30\% during prime time). Netflix has been studied extensively, but not with opportunistic caching performance in mind.

Netflix encrypts the stream manifest (MPD). For each video, the encoding rate is selected by choosing a specific file name. We first needed to map the file names (extracted from the HTTP GET requests) and the corresponding video playback rates, which are not externally visible.
\begin{table}[!t]
  \begin{center}
    \begin{tabular}{lr}
      \multicolumn{2}{c}{Desktop client} \\
      Video playback rate [kbps] & File name \\
      \hline
      235 & 15304768.imsv \\
      \hline
      375 & 22630482.imsv \\
      \hline
      560 & 20693663.ismv \\
      \hline
      750 & 17864652.ismv \\
      \hline
      1050 & 17091067.ismv \\
      \hline
      1750 & 18266773.ismv \\
      \hline
      2350 & 39219201.ismv \\
      \hline
      3000 & 41121933.ismv \\
      \hline \hline
      \multicolumn{2}{c}{Smart-phone client} \\
      \hline
      110 & 14705760.imsv \\
      \hline
      182 & 14986798.imsv \\
      \hline
      257 & 15723334.ismv \\
      \hline
      506 & 15411523.ismv \\
      \bottomrule
    \end{tabular}
    \caption{Mapping of video playback rates corresponding to file names for the movie "The Lorax"}
    \label{tab:videorates}
  \end{center}
\end{table}
The eight different rates between $235$~kbps and $3000$~kbps for the Netflix desktop client, as well as the four different rates ranging from $110$~kbps to $506$~kbps for the Netflix Android smart-phone client can be seen in TABLE~\ref{tab:videorates}, as well as the file names corresponding to these rates for the movie "The Lorax". For monitoring the client-server communication we used \textit{tcpdump} and \textit{tcpstat} for data evaluation.

We first studied a stable and typical Netflix connection: it is a laptop connected through WiFi to a broadband network. The laptop is staying at the same position and the Netflix client has enough bandwidth to sustain the highest available rate. We observed that the rate adaptation mechanism for the video stream oscillates between the highest two rates. The oscillations may be due to variations on the wireless channel (which we did not influence) and to some exploration attempts from the rate adaptation mechanism to verify the performance on other rates. A wide number of measurements at different times showed a similar behavior.

\begin{figure}[!t]
\centering
\includegraphics[width=3.5in]{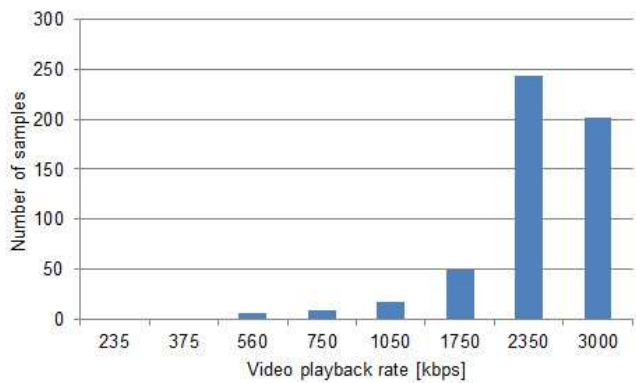}
\caption{Rate distribution of a Netflix desktop client in a stable set-up.}
\label{fig:netflixHist}
\end{figure}

\begin{figure}[!t]
\centering
\includegraphics[width=3.5in]{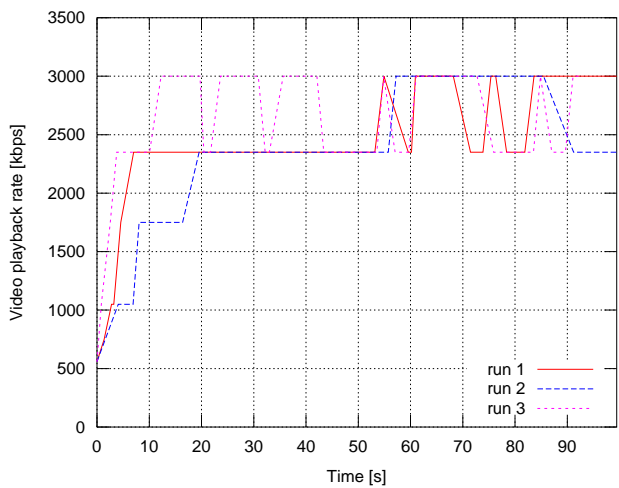}
\caption{Correlation of Netflix clients based on three different runs.}
\label{fig:netflixCorr}
\end{figure}

From the histogram in Fig.~\ref{fig:netflixHist}, one can see that even in a stable environment with no user mobility and excellent wireless channel conditions, the rate selected by the Netflix client after the initial bootstrapping period, converges to an oscillating pattern where it roughly splits between the two highest rates. We can thus model as a two-state Markov chain with equal probability to be in either state. The presence of the other (lower) rates is due to the \textit{slow start} of the Netflix client, which can be seen in the first 10~seconds in Fig.~\ref{fig:netflixCorr}.

We also measured the similarity of the traffic between different successive downloads. For this, we placed the client into a static position, and repeated the streaming of a sequence of a Netflix movie to see whether or not the trajectory followed through the (time,rate) domain would repeat over time.
We found that they are independent in that the selected rates follow the same distribution, but for any point in time, the outcomes become uncorrelated, as shown in Fig.~\ref{fig:netflixCorr} for three independent runs of the same streaming service. In essence, this implies that the efficiency of caching in the network would be diluted by the adaptive streaming mechanism, as subsequent requests for the same video may not overlap in the rate dimension.

We repeated this experiment with a second Netflix client watching the same video.
\begin{figure}[!t]
\centering
\includegraphics[width=3.5in]{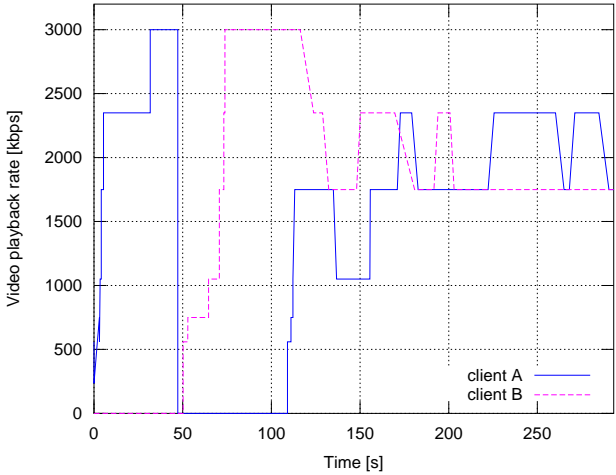}
\caption{Video playback rates used by two competing Netflix clients sharing the same bottleneck.}
\label{fig:twoClients}
\end{figure}
Fig.~\ref{fig:twoClients} shows that both clients are able to pick the highest rate in absence of each other. After about 110~seconds both clients share the bottleneck and the rate reduces. One can see, although some oscillations occur, each client gets its fair share, in other words they have similar (or equal) rates at any point of time. In this case, the evolution of the rates is synchronous: both trajectories would overlap, if both clients sharing the bottleneck are watching the same video.

\begin{figure}[!t]
\centering
\includegraphics[width=3.5in]{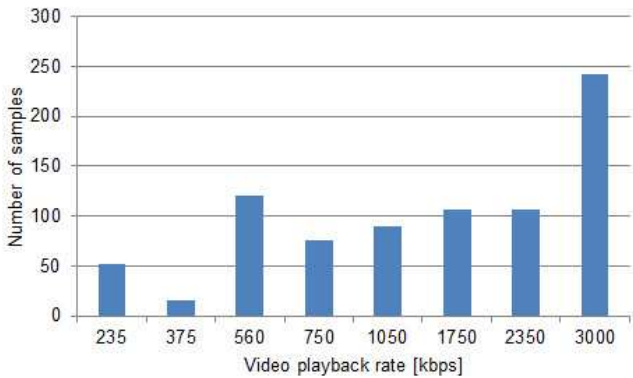}
\caption{Rate distribution of a Netflix client hosted by a mobile notebook.}
\label{fig:realChannelHist}
\end{figure}

Finally, we perform the adaptive streaming experiment with a mobile client: a user moves around while staying within range of one access point. We see the rate distribution in Fig.~\ref{fig:realChannelHist}.
In this case a wider number or even all rates are used because of the highly varying channel conditions. While in good network condition, the video stream was split over two rates. This spread increases when the network conditions deteriorate.

To summarize this empirical evaluation, we see that the dynamic adaptive streaming spreads the packet requests over a wide number of rates, and that streaming a specific video becomes the transfer of a sequence of segments that follow a trajectory over a (time,rate) space which is unlikely to be repeated by other streams of the same video. Even in the best-case scenario of a stable set-up over an excellent quality connection, we still observe an oscillation over two states.

\subsection{Effect of Caching On Rate Adaptation}

We observe another form of oscillation when a web proxy or in-network cache is inserted between the clients and the server. For this, we used a DASH implementation as Netflix makes it difficult to store its video content.

We extended the ICN architecture detailed in \cite{Chanda2013,Chanda2013a} and interfaced it to the open source DASH client~\cite{Muller2011}. We configured the network setup on a single Linux machine: an Open vSwitch connecting four virtual machines (VMs). A DASH client, a proxy, a content server, and a cache reside on the four different VMs, respectively. Links between these network elements show a RTT of $1$~ms for standard ping requests. A Floodlight OpenFlow controller is also running to regulate and modify flows when needed. The DASH client delegates all its HTTP connections to the proxy, which consults the controller for content location upon interception of an HTTP request from the client. In case of a cache miss, the server will respond with the video content, and the controller will ensure that a copy of the content being forwarded to the cache. Otherwise, the cache will return the content to the client.

\begin{figure}[!t]
\centering
\includegraphics[width=3.5in]{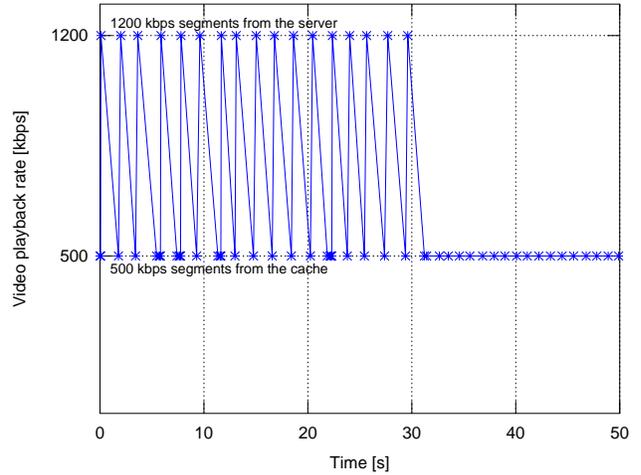}
\caption{Cache-server oscillation.}
\label{fig:cacheOscillation}
\end{figure}

The server VM contains DASH-compatible video datasets~\cite{Lederer2012}. Each video is available as a collection of equal-length small segments and each segment is available at a variety of bit rates. Moreover, each segment at a certain bit rate is named uniquely. Therefore, after the DASH client acquired the MPD file from the server, it becomes aware of the available bit rates for the subsequent video segments and is able to identify the segments from the HTTP request. However, whether the segments are available at the cache is only known at the controller. As illustrated earlier, the oscillation effect can happen
when only the low bit rate segments are being cached, whereas the client-cache link is far better than the client-server one. Specifically, the HTTP requests are being directed to the server and cache back and forth due to the adaptive rate selection policy of DASH. We demonstrate this effect with our prototype by imposing a delay of $100$ms on the client-server link, and Fig.~\ref{fig:cacheOscillation} shows the results.
As it can be seen the rate oscillates between $500$~kbps segments received from the cache, and
$1200$~kbps segments sent by the server. At $t=30$~s, we added a $100$~ms delay also at the client-cache link. Figure~\ref{fig:cacheOscillation} shows that from that time on, the oscillation stops as the client-cache link quality dropped such that the rate adaptation algorithm determines a higher bit rate is no longer sustainable.

We have shown two negative impacts of ICN for DASH-based video streaming: a reduced hit rate at the cache caused by the multiplication of encoding rates; and an imprecise rate estimation caused by an oscillation between the client-cache and the client-server channels. We now analyze the performance of in-network caching theoretically based upon our empirical observation.

\section{Characterizing the Cache Hit Rate of Adaptive Streaming}
\label{sec:characterizing}

Intuitively, the performance of in-network caching using an adaptive streaming mechanism would be affected as follows: consider a cache of capacity $C$. Instead of caching one copy for each item under a non-adaptive mechanism, $C$ would now split its capacity to accommodate multiple rates. For instance, if the two rates used are $3,000$~kbps and $2,350$~kbps, then we would expect roughly three fifth of the cache to be used up by the first rate, and two fifth of the cache used up by the second rate. If the non-adaptive streaming uses only one single rate of $2,675$~kbps (such that the total rate is the same in both non-adaptive and adaptive case), then the cache could hold $C/2,675$ units of traffic, vs $C/(3,000+2,350)$ in the adaptive case. This means the cache size is effectively divided by two with adaptive streaming.

Another intuitive way to look at it is as follows: upon requesting a segment $s_i$, the adaptive stream will pick one rate $s_i^j$. The next request for this segment will be for rate $s_i^l$, and if there are only two rates and both rates are equally likely, the probability that both are for the same rate is $1/2$. So in this approximation, the hit rate is divided by two for two subsequent video streams.

We now try to confirm this intuition by formalizing the problem to compute the hit rate probability in adaptive vs non-adaptive streaming. We build a simple model to analyze the performance of ICN architectures which support in-network caching. We assume that we have $M$ objects (say, videos) each composed of $S$ segments. For simplicity of the analysis, we suppose that each object has the same number of segments, and each segment at a given encoding has the same size. Based on our empirical analysis, we consider for now that only $R = 2$ encoding rates are used, $r_1$ and $r_2$.  At the server, the object $X$ will contain all the possible time segments and encoding rates, that is $X = \{s_i^j, i = 1,\ldots, S; j = 1,\dots, R\}$ where $i$ is the time index, and $j$ is the rate index. We denote by $b_1$ and $b_2$ the size of the segments for the two encodings. For instance, in our empirical observation, $b_1 = 3,000$~kb and $b_2 = 2,350$~kb.

Let's assume that an object is requested with a popularity distribution drawn from a Zipf-like distribution with parameter $\alpha$, namely that the probability $p_k$ of the $k$-th item to be requested is proportional to $1/k^\alpha$. $\alpha$ is typically between $0.5$ and $1.5$ for most Internet traffic (for instance,~\cite{Fayazbakhsh2013} finds a value near $1$ and~\cite{Wolman1999} near $0.7$). We suppose further a queue which sees the requests from the client, and caches the traffic opportunistically as it goes through. We assume that this cache operates under a LRU policy (studied in~\cite{Asit1990} for instance). We also make the simplifying assumption that all time segments are requested with the same probability, i.e. that there is no attrition during the course of the stream (this is not true, but \cite{Roberts2013} confirms that it is a reasonable assumption to make due to the large number of items). Finally, we suppose the rates are equiprobable, namely that $s_i^j$ is requested with the same probability independently of $j$.

The probability $p(s_i^j)$ of a specific segment $s_i^j$ being requested from the $k$-th item is:
\begin{eqnarray}
p(s_i^j) & = & \frac{ \frac{1}{SRk^\alpha}}{\sum_{u=1}^{u=M} \sum_{v=1}^{v=S} \sum_{w=1}^{w=R} \frac{1}{SRu^\alpha}} \nonumber\\
& = & \frac{p_k}{SR}
\end{eqnarray}

Under the Che approximation for LRU queues~\cite{Che2002} 
the probability $\pi_{k,i,j}$ that the $i$-th segment with coding $j$ of object $k$ is in the cache of size $C$ is:
\begin{eqnarray}
\pi_{k,i,j} = 1 - e^{-\frac{p_k}{SR} t_C}
\end{eqnarray}
where $t_C$ is the unique number such that:
\begin{eqnarray}
C = \sum_{k=1}^M \sum_{i=1}^S \sum_{j=1}^R b_j \pi_{k,i,j}
\end{eqnarray}

In our case,
\begin{eqnarray}
C = S \left(\sum_{j=1}^R b_j\right) \left( \sum_{k=1}^M (1 - e^{-\frac{p_k}{SR} t_C} ) \right)
\end{eqnarray}

The cache hit probability $P(hit)$ is therefore:
\begin{eqnarray}
P(hit) & = & \sum_{k,i,j} \frac{p_k}{SR} \pi_{k,i,j} \nonumber \\
       & = & \sum_k p_k (1 - e^{-\frac{p_k}{SR} t_C})
\end{eqnarray}

We can now compute the probability $P(hit)$ for different $R$. In particular, for $R = 1$, we would have a non-adaptive streaming where all the clients request the same encoding rate $r_1$. We can therefore compare the performance of opportunistically caching in the network under a non-adaptive streaming (which is the typical assumption in ICN) and an adaptive streaming using $R > 1$ rates.

\begin{figure}[!t]
\centering
\includegraphics[width=3.5in]{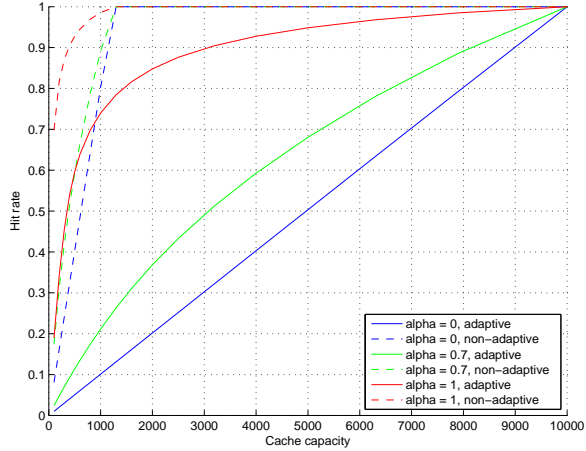}
\caption{Cache hit probability for non-adaptive vs adaptive streaming.}
\label{fig:hitRate}
\end{figure}

In Fig.~\ref{fig:hitRate} the cache hit rate versus the the cache size for different values of $\alpha$ is plotted.
We have set $R = 8$ for adaptive streaming. It can be seen that for uniform distribution ($\alpha = 0$), the hit rate is effectively divided by a factor of eight, comparing the adaptive with the non-adaptive case. Also, the higher $\alpha$, the more the distribution of requests is concentrated over only a few items, and the hit rate in this case is not significantly affected. This means that the most popular items are requested often enough that all rates end up in the LRU cache. However, in the Internet, a typical value of $\alpha$ is $0.7$, which results in a significant degradation of the in-network caching performance. For instance, for a cache size of $1000$ and $\alpha = 0.7$, we see that the cache hit rates degrades about 75\%.

The rate request process will depend on the network performance seen by the nodes, and the analysis above can be combined with the typical throughput distribution seen in, say, cellular networks~\cite{Mukherjee2012} to calculate the hit rate of a cache placed at the edge of such networks.

\section{Cache Oscillations}
\label{sec:cacheOscillation}

We now consider the throughput oscillations observed at the streaming client that are induced by the presence of the cache.

Let's assume that the throughput from the cache is $\omega_c$ while the throughput for the server is $\omega_s$. The throughput in the cache is higher due to the shorter round-trip time and to the underlying behavior of TCP. Further, the available bandwidth between the client and the server will be the minimum of the minimal bandwidth between the client and the on-path cache, and the cache and the server. The ratio of theses two throughputs is denoted by $\rho = \omega_s / \omega_c$.

Further, we assume that the available rates are separated by a factor $\rho_r$. For numerical purpose, we take $\rho_r = 1.5$, namely $r_2 = 1.5 r_1$, $r_3 = 1.5 r_2$, and so forth. This is roughly the ratio of contiguous rates used by the Netflix desktop client (the actual ratios are $1.60, 1.49, 1.34, 1.40, 1.67, 1.34, 1.28$).

We consider the probability of hitting a segment in the cache to be $\pi_k = (1 - e^{-\frac{p_k}{SR} t_C})$ (we remove the dependency on $i$ and $j$ from the index since the probability is uniform across all $i$'s and $j$'s). We assume that the rate adaptation mechanism bases its rate evaluation upon receiving $q$ segments. The mechanism functions this way: the rate is set based on the performance of the download of the last $q$ segments. For each segment, the time to complete the download of the segment is estimated. This is averaged over the last $q$ segments to yield the throughput between the content provider and the content consumer.

The estimated available rate at the end of the observation window $T$, $\omega(T)$, is considered to be:
\begin{eqnarray}
\omega(T) = \frac{1}{q} \sum_{t=1}^q \frac{b(T-t)}{\tau(T-t)}
\end{eqnarray}
where $\tau(t)$ is the time to download the $t$-th segment.

$b(T-t)$ is constant over the $q$ segments and the rate is only re-assessed at the end of the $q$ segment window. Therefore the ratio will be equal to $\omega_s$ if the segment is fetched from the server, or $\omega_c$ if it is fetched from the cache. We need to discuss two cases: for small $q$ and for large $q$.

For high values of $q$, then we can approximate $\omega(T)$ with:

\begin{eqnarray}
\omega(T) & = & \frac{1}{q} \sum_{t=1}^q \left( (1 - e^{-\frac{p_k}{SR}t_C}) \omega_c + e^{-\frac{p_k}{SR}t_C} \omega_s \right) \nonumber \\
& = &  \left(1 + e^{-\frac{p_k}{SR}t_C} (\rho - 1) \right) \omega_c
\label{eq:largeq}
\end{eqnarray}

Note that this becomes deterministic, namely that the video streaming rate will be able to estimate the performance of a joint channel which combines cache hits and service hits (upon cache misses). By requesting the encoding corresponding to this average rate, the streaming mechanism is able to become independent of the content location.

For low values of $q$, $\omega(T)$ becomes:
\begin{eqnarray}
\omega(T) =  \frac{1}{q} \sum_{t=1}^q \left( \mathbf{1}_{\{s_t^j \in \mbox{ Cache }\}} \omega_c  + \mathbf{1}_{\{s_t^j \notin \mbox{ Cache }\}} \omega_s \right)
\end{eqnarray}

Since $\mathbf{1}_{\{s_t^j \in \mbox{ Cache }\}}$ is a Bernoulli random variable with parameter $(1 - e^{-\frac{p_k}{SR} t_C})$, we can  calculate the distribution of $\omega(T)$.

Note that the actual rate used by the connection will be $\omega_c$ or $\omega_s$. In particular, $\omega(T)$ is equal to the desired rate if and only if the last $q$ segments and the next segments are all fetched from the same location (that is: all cache hits or all cache misses).
In all other cases, the rate achieved is suboptimal: either to high and the user might see degraded performance, or too low and some capacity was left unused.

\section{DASH-INC: A Cache-Aware Adaptive Streaming}
\label{sec:guidelines}

Based upon our experiments and analysis, we propose an ICN-aware dynamic adaptive streaming.

\subsection{Cache-Server Oscillations}

To solve the cache-server oscillations, one obvious fix is to increase the observation window $q$ (see equation~\ref{eq:largeq}). For large $q$, the estimated rate at the client includes a mix of packets served from the cache and from the server. This would smooth out the rate changes and the client would request a rate that is adapted to the hit rate at the cache.

Another potential fix would be to set the rate transitions at some predefined intervals. For instance, the rate transitions could occur only when the segment index is equal to 0 modulo $q$. This would restrict the oscillations as the rate could not ping-pong between the cache and the server until the next allowable transition. This would also increase the cache rate, since it would synchronize the rate transitions among all the nodes. Therefore two nodes seeing similar conditions would end up following similar trajectories. Consider the three subsequent streams of Fig.~\ref{fig:netflixCorr}: they would transition to the same rate at the same time, and have a much higher hit rate at the cache.

However, both suggestions would be at the expense of reactivity to the network conditions (wireless channel, network congestion, etc). For this reason, we propose below some ICN-specific improvements.

\subsection{DASH-INC}

We now describe DASH-INC, an adaptive streaming mechanism for in-network caching (INC). We assume an ICN-like architecture, where the client requests a segment by sending an interest (that is, there is a GET primitive) and any node that can respond to this interest is allowed to do so (namely, there is late binding of content to location). We assume our ICN architectures still has some notion of location, so that the client can tell apart content hosted at the server from content hosted at the cache. \cite{Chanda2013} describes such an architecture. The content is uniquely identified, such that a specific rate and time segment can be mapped to its stream based upon its name only (unlike the Netflix format described in Table~\ref{tab:videorates}, which assigns a different name to each rate). For instance, all segments of a given stream can share a common, unique name prefix.

\begin{figure}[!t]
\centering
\includegraphics[width=3.5in]{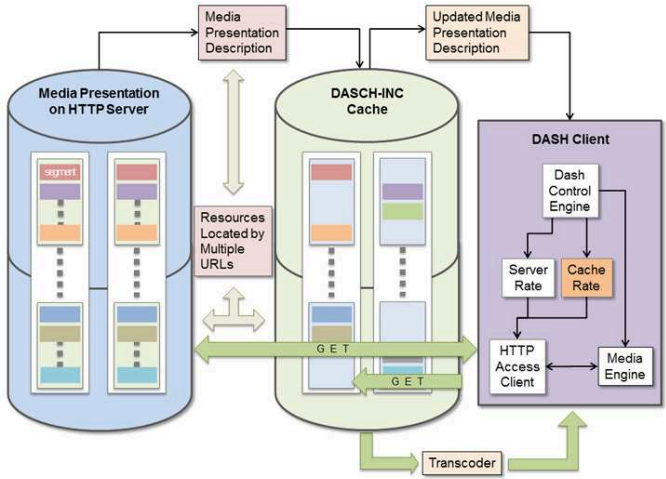}
\caption{Dash-INC architecture.}
\label{fig:dashInc}
\end{figure}
We propose an ICN-based DASH architecture, as it can be seen in Fig.~\ref{fig:dashInc}, with the goal of improving the performance at the client by using the cache as much as possible (lower delay, better QoE for the end user, less peering traffic for the network provider) while at the same time limiting the cache-server oscillations.

\subsubsection{MPD Modification}

When the client starts a DASH video stream, it requests first the MPD manifest. The MPD includes a list of servers and different encoding and rates. In DASH-INC, when the client sends an interest for the MPD, the intermediate router that holds the content can respond in two ways: if it has cached the MPD previously, it can send the manifest back without forwarding the interest to the origin server. However, it includes itself, and the available rates that it has cached into the MPD\footnote{We do not consider here the security issues associated with intermediate nodes modifying the content; the cache can sign the content using its own keys and deliver the manifest as its own. DRM (digital rights management) issues and security issues are beyond the scope of this document.}. A brute force description of each rate and time segment available at the cache can be included, but we suggest using some compression mechanism to keep the MPD size manageable. Note that the MPD can be split into several different manifests in order to keep the MPD of relatively small size.

The client then receives an MPD which points to several potential locations each with different available segments. The rate adaptation mechanism at the client must now be modified as part of the client, so that the client keeps track of the achieved rate for each content source. Namely, the client requests a GET that specifies the preferred rate. It receives the corresponding packet. However, this packet can come from a cache or the source. The client keeps track of the provenance of the packet, and updates only the corresponding rate (say, rate for cache-client channel, or rate for server-client channel).

The client then picks the highest available rate that is supported based upon the rate calculation.

\subsection{Cache Modification}

In DASH-INC, the cache can modify the MPD to alert the client of the available rates. However, we have seen in section~\ref{sec:characterizing} that the multiplication of encoding rates negatively impacts the cache performance.

We modify the cache in the following manner: upon receiving a packet, the cache checks if it holds a packet corresponding to the same video stream, same time segment, but different encoding rate. If it does, then it keeps only the highest rate.

Upon receiving a request for a lowest rate than the copy held in the cache, the cache then transcodes the highest rate into the lowest rate. This requires to equip the cache with a transcoding mechanism. However, those are commonplace nowadays and can transcode video streams at line speed\footnote{Companies like HeyWatch, Encoder.com or Zencoder perform transcoding as a cloud based service.}. The cache also refreshes the cache content by placing the highest-rate copy back to the top of the cache (in a LRU policy).

For popular content, this ensures that the cache will end up populated with the highest rate copy, and that from this copy, it is able to generate all other requests. This reduces the cache usage (a space of $3,000$~kb is required for each time segment rather than $10,070$~kb for the Netflix rates) and therefore increases the cache hit rate by making space in the cache for more packets, and having an answer for any request for any rates for packets held in the cache.

For a video stream that is cached entirely at the highest rate, the cache does not have to modify the MPD any longer, as it can answer any request.

\section{Discussion \& Conclusions}
\label{sec:conclusions}
In this paper we looked at two trends affecting online video streaming: a convergence of most content delivery platforms towards dynamic video streaming over HTTP, and caching at intermediate points within the network architecture.

Even if a static client experiences a stable WiFi connection, providing enough bandwidth to achieve the highest video quality, the video stream still oscillates between the two highest rates, in a uniformly distributed manner. For non-static clients an even wider number of rates are used because of highly varying channel conditions. Although the outcomes for different successive downloads look similar the chosen rates are not correlated. Our observations lead to the observation that the transfer of a sequence of segments that follow a trajectory over a (time,rate) space is unlikely to be repeated by other streams.

Based on our results we estimate the benefits of in-network caching. Using adaptive streaming the effective cache size is divided by the number of different representations stored in the cache. Storing only one representation, respectively the segments with highest quality, and derive all other representations by \textit{on the fly transcoding} can overcome this excessive use of memory at the cache.

There is a trade-off between the complexity of processing content and the cache hit rate: the cache hit rate is increased by the addition of a transcoding element. The server typically caches the different rates to reduce the processing complexity: since storage price is decreasing faster than CPU, it is more cost-efficient to add storage at the origin server (or at the CDN) rather than adding processing power. At the cache, the calculus is different: the server knows how much storage is required and can be dimensioned accordingly. However, for an in-network cache, there is no limit to the amount of storage that is required due to the fact that the client can request any file, from any server.

By inserting copies of video segments at points within the client-server path (caches) the channel is modified. Therefore the rate adaptation algorithm (based on throughput estimation) observes (usually) better channel conditions which can lead to cache-server oscillations and poor allocation of resources. Instead of only observing the currently used connection both paths (client-cache plus client-server) should be estimated by the client separately. A rate adaptation in context of requesting segments at the server should only take place if the available throughput is sufficient.

We have described DASH-INC, a mechanism to solve the issues due to the interactions of adaptive video streaming with in-network caching. An implementation of DASH-INC in our testbed is left for further study.

\bibliographystyle{IEEEtran}
\bibliography{library}

\end{document}